# Identification and Classification of Phenomena in Multispectral Satellite Imagery Using a New Image Smoother Method and its Applications in Environmental Remote Sensing

M. Kiani[a]

[a] *department of surveying and geospatial data engineering, university of Tehran, Iran*

**Abstract**

In this paper a new method of image smoothing for satellite imagery and its applications in environmental remote sensing are presented. This method is based on the global gradient minimization over the whole image. With respect to the image discrete identity, the continuous minimization problem is discretized. Using the finite difference numerical method of differentiation, a simple yet efficient 5 × 5-pixel template is derived. Convolution of the derived template with the image in different bands results in the discrimination of various image elements. This method is extremely fast, besides being highly precise. A case study is presented for the northern Iran, covering parts of the Caspian Sea. Comparison of the method with the usual Laplacian template reveals that it is more capable of distinguishing phenomena in the image.

## 1. INTRODUCTION

There has always been the need to monitor the earth's surface in order to have a better understanding of various phenomena that occur on it, particularly those that have a great and immediate impact on the humanity's life. Various ways have traditionally existed, including ordinary, simple land surveying method. However, as the humans evolve and their lives become complicated, there is need for better understanding of the matrix-the earth-on the surface of which they live. In the modern era of technology, in particular, this need has increased drastically. From the anthropogenic changes in the landscapes, loss of valuable biodiversity, and the natural trends and hazards, to the estimation of crop yield and the fate of the chemicals that are released at an alarming rate to the oceans and seas.

An excellent and powerful response to this growing need for more accurate techniques of measuring and monitoring the mentioned problems was the introduction of space-based techniques, among which are the GNSS, VLBI, and remote sensing satellite imagery. Remote sensing is of great interest because of its accuracy, compared to its relatively low cost of obtaining the product-image. In this paper, we focus on remote sensing and its applications in environmental monitoring.

Various methods and techniques are employed to implement analyses on the satellite imagery. From the image enhancement and object identification to classification and producing byproducts. One especially important aspect of the remote sensing procedure is the identification of various elements in the image. This is done in a variety of ways and is pursued by various authors. [1]-[8], in particular, have contributed to the methods of distinguishing phenomena in the image. Sometimes, even an anomalous thing can be detected in the image, by the so-called methods of anomaly detection.

In [1], a new method of image smoothing is presented, called optimal image smoothing. The norm of the Laplacian operator is minimized, which results in the biharmonic (second degree Laplacian) equation. The minimization problem is discretized and a 5×5 template is derived. The application of the propsed method is presented in the Landsat multispectral satellite imagery, whereby the anomalies in the Qom region in central part of Iran are detected. Particularly, the famous Salt-lake is identified. In [2], the first physiologically motivated object detection based on the neural network method is presented. The primates' system of vision and its characteristics such as expectation and temporal synchronization are the bases on which the method works. The method presented in this article is primarily used alongside the other object detection methods, to improve the accuracy of the object detection procedure. This means that the usual object detection techniques in the image are applied and then the role of the new method is to combine different elements to decide what the object is. To demonstrate the accuracy of applying this new method to object detection procedure, the mammograms and forward-looking infrared radar images are used. It is shown that the method is more accurate than the usual methods used for the purpose of object detection. In [3], a new method is presented for object detection in the VHR images. This method is based on convolutional neural networks. This method arranges different layers of information, in which the lower levels contain valuable details and the high levels valuable semantic information. Also, another layer is defined, which combines the information in low and high levels. It is shown that the proposed method is much more accurate than the rest of the available methods. In [4], a new method for object detection in serial images is presented. In this method the Gauss statistical properties for deriving a background, based on which the object is detected by subtracting the built image from the moving image, plays the paramount role. Also, the accuracy of the object detection is improved, with respect to the other methods. In [5], unlike the usual, prevalent low-level object detection methods, which perform the color-similarity in two-dimensional image space, the technique of detecting objects using the edge color distribution space is employed. The three-dimensional case is also studied, and a new edge-tracking algorithm is presented. The accuracy of the methods presented in this paper is good, which is shown through the experiments on various images. In [6], a new method of object detection is presented for the high-resolution SAR satellite imagery. If the background around the object is complex, then the traditional SAR images object detection methods are not of much use, since their accuracy levels are not high enough. However, the proposed method in this paper, which is based on the fundamental characteristics of the optical imagery, improves the accuracy of object detection algorithm. As an indication of the accuracy improvement, the problem of inshore ship detection is presented. Since in this particular problem the surrounding ambience of the object is relatively complex, the proposed method works much better, compared to the usual object detection algorithms. In [7], a new method is presented for object detection in remote sensing video. Since the small objects in images taken by optical sensors of the satellites are usually blurred, this new technique uses the concept of weak object detection. The proposed method is based on combining the temporal information from the neighboring frames and the spatial information in the image itself. Comparison of the results of applying this new method and the other methods reveals that it improves the accuracy of detecting moving objects. In [8], a new method of object detection is proposed, which uses the concept of object localization and object classification. Since the localization and classification need two different layers of information, the usual object detection algorithms perform these in two different phases. However, the novelty of the proposed method is that it unifies the two phases, performing them jointly in one phase. The real data analysis shows that this method excels the existing methods.

Thus, there exist a variety of methods for object detection and classification. The need of detecting various elements in remote sensing satellite imagery is evident, especially when it comes to the application in environmental remote sensing. Therefore, We are motivated to present a new method of image smoothing, whereby the objects, elements and phenomena in the image are distinguished. This is done using the minimization of the gradient operator over the whole image. Gradient is a measure of change of a function, and its minimization results in a function with maximum degree of smoothness, in the sense that its variation from one cell to another is as small as possible.

The rest of this paper is organized as follows. In section 2, the mathematical formulae for the derivation of the image smoother are presented and the image smoothing template is derived. In section 3, the application of the derived results in section 2 is presented in the field of remote sensing, for the Landsat satellite imagery of the northern Iran. In section 4, a comparison analysis is presented for the new method, in which the rigor of the proposed method is challenged against the established method of Laplacian template. Finally, the conclusions are mentioned in section 5.

## 2. Image smoothing template derivation

In this section, the mathematical formulae for the derivation of the image smoother are presented and the image smoothing template is derived.

The essential mathematical background for the method is the theory of norm minimization in [1] and [9]. Since the gradient is a measure of how fast a function changes, the minimization of the norm of the gradient results in the smoothest function with minimal changes. Hence, in a function with high degree of variability, the smoothest function can be of great use, since with the comparison of this function with the highly variable function one can detect the deviations from the criterion. Therefore, the method presented in this paper is based on the minimization of the norm of the gradient operator. The coordinate system we have used is the usual Cartesian one, denoted by (x,y). To perform the norm minimization, we need to use the gradient operator as the following

$$\nabla = \frac{\partial}{\partial x} + \frac{\partial}{\partial y}. \tag{1}$$

The minimization of the norm of the gradient operator results in a function with utmost smoothness, in terms of differentiation. The definition of the minimization problem is given in the following.

***Definition 1:*** The function $f$ is the solution of the gradient minimization problem if it satisfies the following relation

$$I(f) = \iint |\nabla f|^2 dxdy \to min. \tag{2}$$

Note that based on (2), the function $f$ must be at least one time continuously differentiable. However, as we will see in the next section, it must be at least two times continuously differentiable.

*2.1. The solution of the minimization problem*

The minimization problem in (2) has a unique solution. In fact, based on the method presented in [9], and using the method of Euler-Lagrange minimization, we can write

$$dI(f) = \lim_{\lambda \to 0} \frac{I(f + \lambda g) - I(f)}{\lambda} = 0, \tag{3}$$

where $g$ is another two-time differentiable function.
In (3), the limit can be computed as follows

$$I(f + \lambda g) = \iint |\nabla f|^2 dxdy + 2\lambda \iint |\nabla f \nabla g| dxdy + \lambda^2 \iint |\nabla g|^2 dxdy. \tag{4}$$

$$dI(f) = \frac{1}{\lambda}\big(I(f + \lambda g) - I(f)\big) = 2 \iint |\nabla f \nabla g| dxdy. \tag{5}$$

Hence, based on (5) and (3) we have

$$\iint |\nabla f \nabla g| dxdy = 0. \tag{6}$$

Regarding [9], we need the eigenvalues and eigenfunctions of the gradient operator. However, it is nice to notice that we do not need the explicit form of these values and functions. Rather, we need to symbolize and use them in the derivation of the image smoothing template. The definition of the eigenvalue and eigenfunction for the gradient operator is given in the following.

***Definition 2:*** For first order differential operator $\nabla$ the m'th eigenvalue, $l_m$, and the m'th eigenfunctions, $e_m$, are the solution of the following differential equation

$$\nabla e_m = l_m e_m. \tag{7}$$

***Theorem 1:*** The solution of the minimization problem (i.e. function $f$) holds in the following second-order differential equation

$$\nabla^2 f = 0, \tag{8}$$

where $\nabla^2$ denotes the Laplacian operator.

***Proof:*** The proof is similar to what is stated in [9]. Hence, based on (6) and considering the orthogonal expansions of the functions $f$ and $g$ based on the orthonormal eigenfunctions, we can write

$$f(x,y) = \sum_{p,q} u_{p,q} e_{p,q}, \tag{9}$$

$$g(x,y) = \sum_{r,s} v_{r,s} e_{r,s}, \tag{10}$$

$$0 = \iint |\nabla f \nabla g| dx dy = \iint \left| \nabla \sum_{p,q} u_{p,q} e_{p,q} \ \nabla \sum_{r,s} v_{r,s} e_{r,s} \right| dx dy, \tag{11}$$

$$= \iint \left| \sum_{p,q} l_p u_{p,q} e_{p,q} \sum_{r,s} l_r v_{r,s} e_{r,s} \right| dx dy, \tag{12}$$

$$= \sum_{p,q} l_p^2 u_{p,q} v_{p,q}, \tag{13}$$

$$= \iint \left| \sum_{p,q} l_p^2 u_{p,q} e_{p,q} \sum_{r,s} v_{r,s} e_{r,s} \right| dx dy, \tag{14}$$

$$= \iint |\nabla^2 f| \, |g| dx dy. \tag{15}$$

The equation (15) holds for every $g$. Thus, the only choice that remains for the function $f$ is that it must satisfy the equation given in (8).

Regarding the equation given in (8), we can write

$$\nabla^2 f = \nabla(\nabla f). \tag{16}$$

Using the relation in (1), one can simply rewrite (16) as the following

$$\nabla^2 f = \nabla \left( \frac{\partial f}{\partial x} + \frac{\partial f}{\partial y} \right) = \frac{\partial^2 f}{\partial x^2} + 2 \frac{\partial^2 f}{\partial x \partial y} + \frac{\partial^2 f}{\partial y^2}. \tag{17}$$

Now, to derive the image smoothing template, one needs to discretize (17). This means we must numerically differentiate (17), instead of continuous partial differentiation. In fact, the two-dimensional function $f(x,y)$ must be discretized. So, we denote hereafter $f(x_n, y_m)$ as the value of function at the cell $(x_n, y_m)$. With this notation, and using the theory of finite differentiation in [10], we have

$$\frac{\partial f}{\partial x} = f(x_n, y_m) - f(x_{n-1}, y_m), \qquad (18)$$

$$\frac{\partial f}{\partial y} = f(x_n, y_m) - f(x_n, y_{m-1}), \qquad (19)$$

$$\frac{\partial^2 f}{\partial x^2} = f(x_n, y_m) - 2f(x_{n-1}, y_m) + f(x_{n-2}, y_m), \qquad (20)$$

$$\frac{\partial^2 f}{\partial y^2} = f(x_n, y_m) - 2f(x_n, y_{m-1}) + f(x_n, y_{m-2}), \qquad (21)$$

$$\frac{\partial^2 f}{\partial x \partial y} = f(x_n, y_m) - f(x_{n-1}, y_m) - f(x_n, y_{m-1}) + f(x_{n-1}, y_{m-1}), \qquad (22)$$

Putting the relations (18)-(22) in (17) one gets

$$\nabla^2 f = 4f(x_n, y_m) - 4f(x_{n-1}, y_m) - 4f(x_n, y_{m-1}) + 2f(x_{n-1}, y_{m-1}) + f(x_{n-2}, y_m) + f(x_n, y_{m-2}), \qquad (23)$$

Hence, using the equation (8), one can derive the following template

|   |   |   |   |   |
|---|---|---|---|---|
|   |   |   |   |   |
|   |   |   |   |   |
| **1** | **-4** | **4** |   |   |
| **0** | **2** | **-4** |   |   |
| **0** | **0** | **1** |   |   |

Fig. 1. Incomplete image smoothing template; the central cell is shown in red

However, the template shown in figure 1 is not complete. In order to derive a symmetric template, we can replicate the template in all other three directions, deriving the final template as the following

| | | | | |
|---|---|---|---|---|
| 0 | 0 | 1 | 0 | 0 |
| 0 | 2 | -4 | 2 | 0 |
| 1 | -4 | **4** | -4 | 1 |
| 0 | 2 | -4 | 2 | 0 |
| 0 | 0 | 1 | 0 | 0 |

Fig. 2. The complete, 5 × 5 image smoothing template; the central cell is shown in red

The template shown in Fig. 2 is the basis of the application we will present in the next section. Note that the sum of the elements in the template is zero. This is typical of all the templates used for finding the difference between adjacent cells. In fact, the difference between the nearest cells and the central cell is computed and the differences between the central cell and its surrounding cells are detected. The next levels of cells are diagonals-having the value 2-and the second-distance horizontals and verticals-having the value 1. This means in these levels the image is smoothed. If no difference is detected between the central and the neighboring cells, the returned value will be zero, as the sum of the elements implies. This template has the structure of a thirteen-point difference star. This template is ideal in determining the differences between the adjacent cells.

**1. Application in the field of environmental remote sensing: distinguishing phenomena in the satellite imagery**

In this section an application of the derived template is presented. The field of remote sensing is based on the satellite imagery, a discrete quantity of great value in discerning the phenomena in an area, enabling humans to track, monitor, and possibly control and change the trends in the area of interest. Among the most important applications of the remote sensing is the identification and classification of different phenomena. This is done using the image processing techniques. The template shown in Fig. 2 is in fact an image object discriminator. Hence, it is a useful means to discriminate the image elements. In order to show the efficiency of the presented method, we use Landsat satellite imagery of the northern Iran, covering parts of the Caspian Sea. This satellite imagery is in 7 bands, and in the following figure we have shown the image with the Optimum Index Factor (OIF), which here is represented by bands 1, 4, and 5.

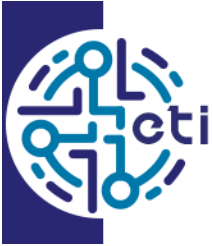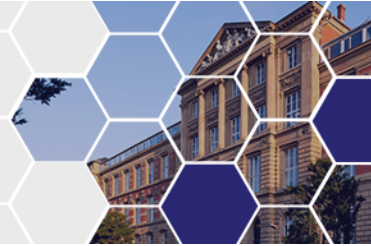

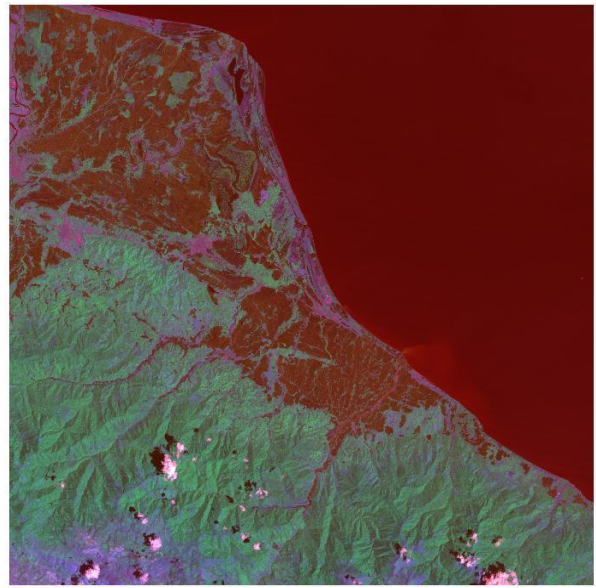

Fig. 3. OIF Landsat satellite imagery of northern Iran and Caspian Sea

The result of applying the template in Fig. 2 to all the 7 bands is shown in Fig. 4 to Fig. 10.

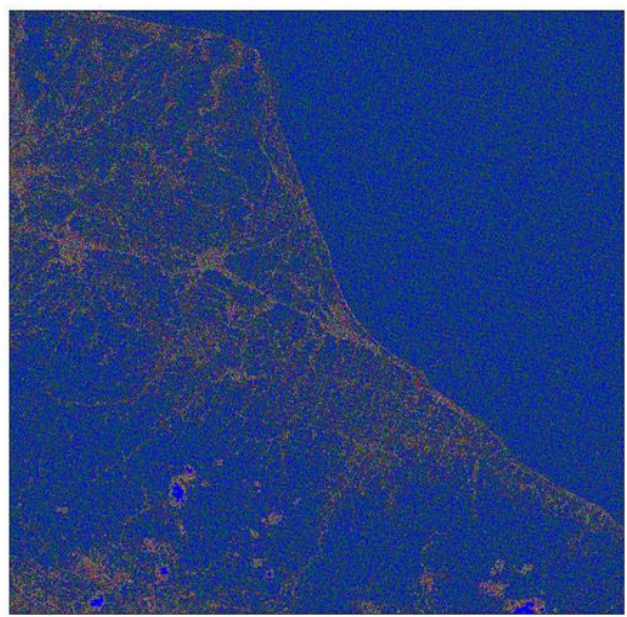

Fig. 4. Different phenomena in band 1

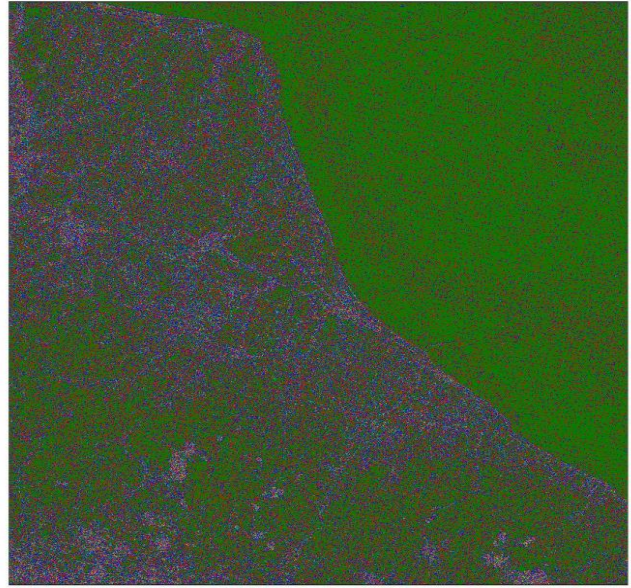

Fig. 5. Different phenomena in band 2

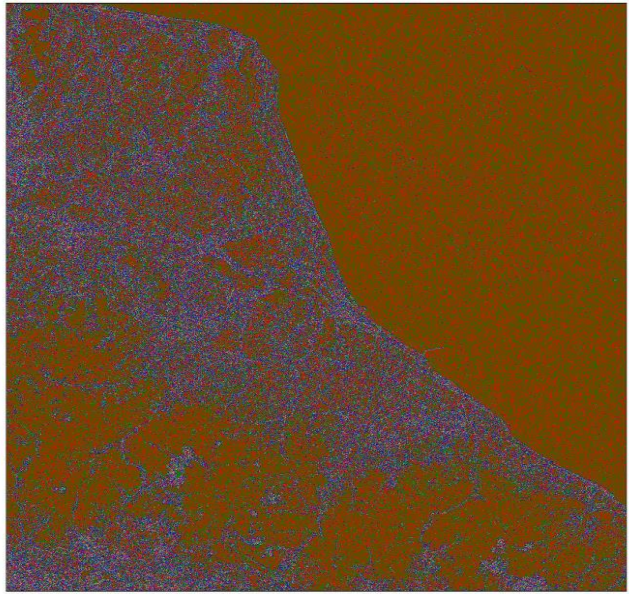

Fig. 6. Different phenomena in band 3

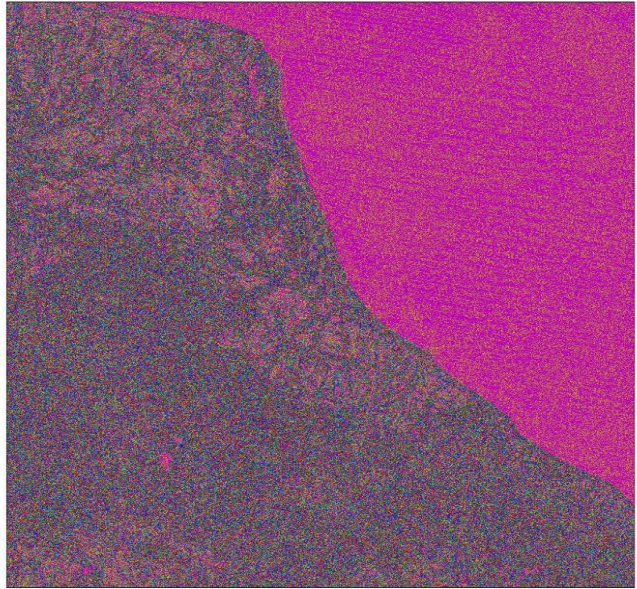

Fig. 7. Different phenomena in band 4

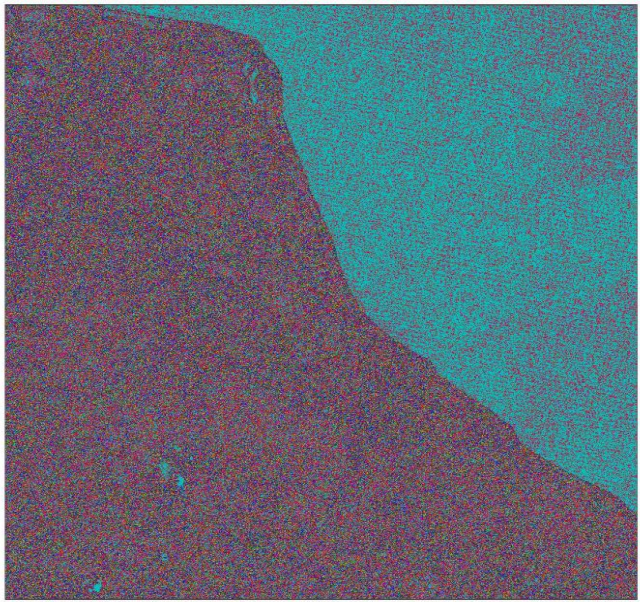

Fig. 8. Different phenomena in band 5

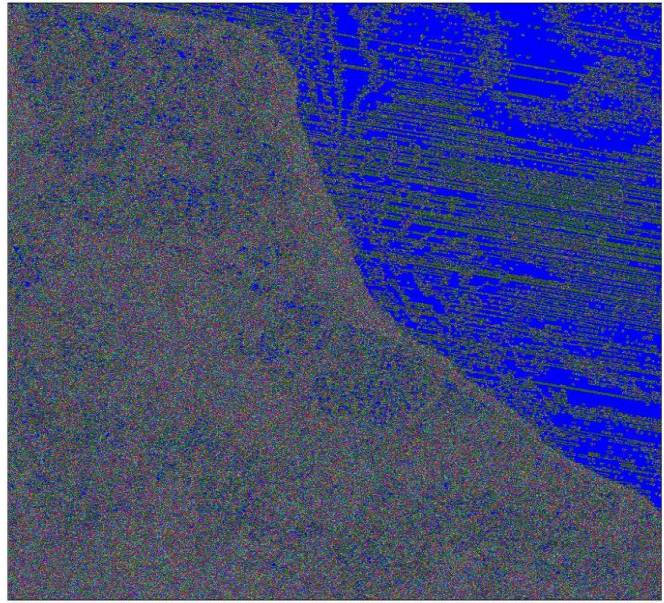

Fig. 9. Different phenomena in band 6

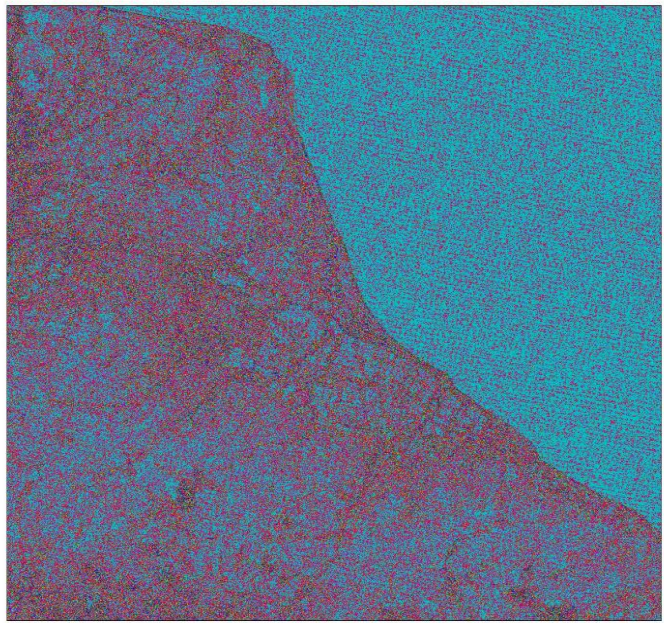

Fig. 10. Different phenomena in band 7

## 2. Comparison with other methods

As one can understand from Fig. 4-10, the Caspian Sea, the jungles, and the soil areas are distinguihsed by the proposed method. The river basins are also discriminated. These can be of great importance in determining the conditions, monitoring, controlling and possibly making policies in the environment.
In this section a comparison is performed between the derived template and the usual Laplacian template. The Laplacian template is shown in the following figure.

| -1 | -1 | -1 |
|----|----|----|
| -1 | 8  | -1 |
| -1 | -1 | -1 |

Fig. 11. Laplacian template

The result of convoluting the Laplacian template with the mentioned satellite imagery in 7 bands is as follows.

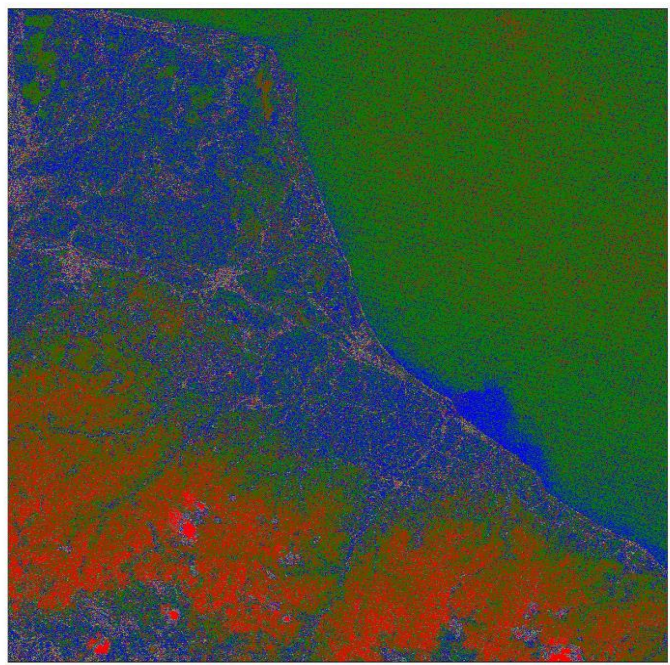

Fig. 12. Different elements in band 1, distinguished by Laplacian template

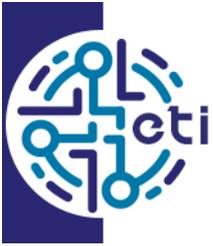
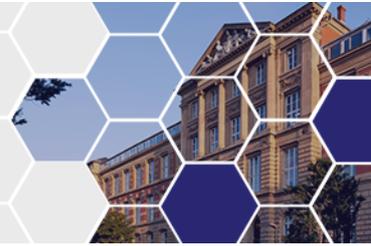

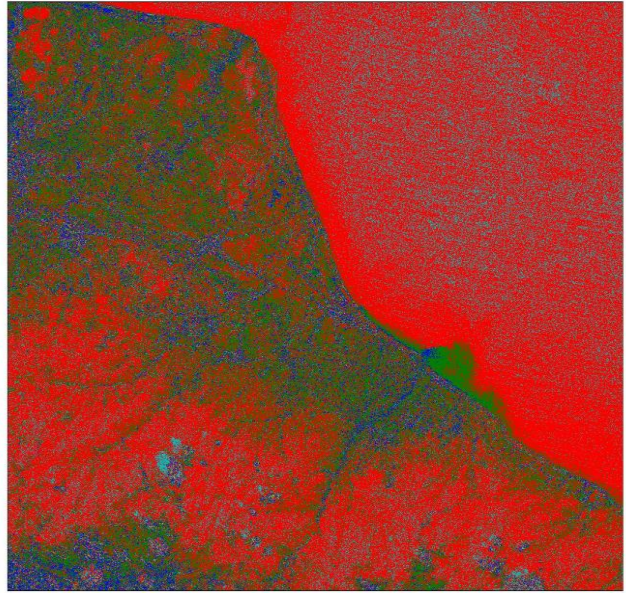

Fig. 13. Different elements in band 2, distinguished by Laplacian template

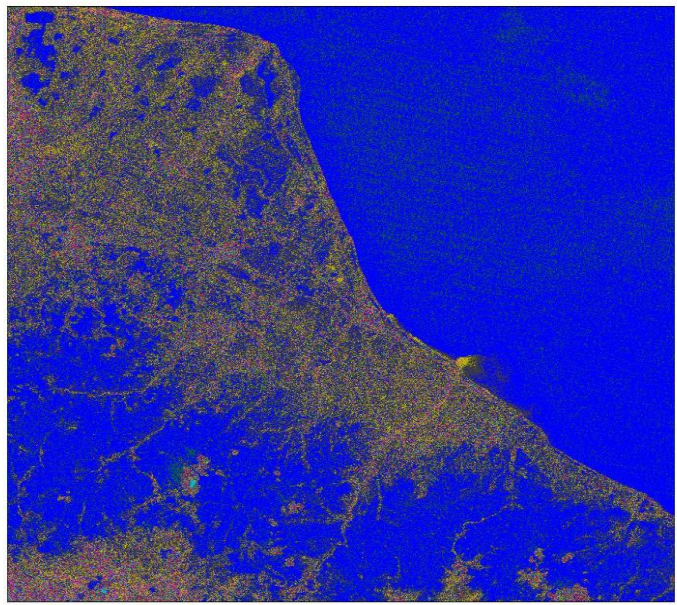

Fig. 14. Different elements in band 3, distinguished by Laplacian template

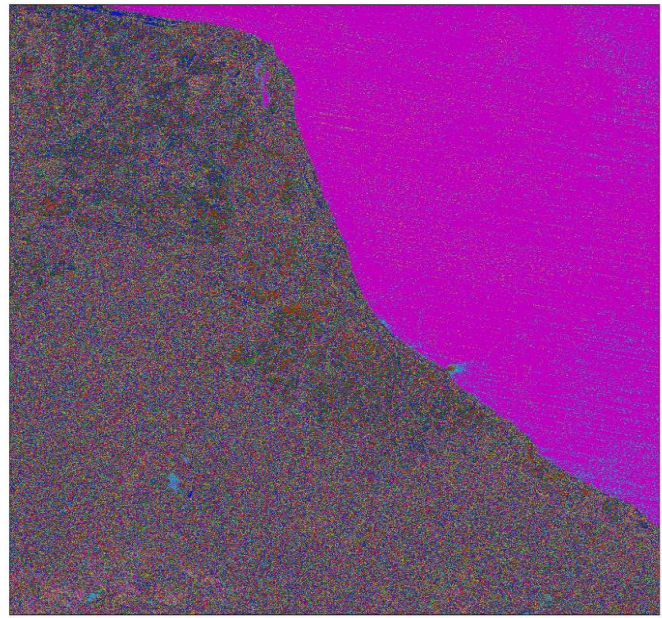

Fig. 15. Different elements in band 4, distinguished by Laplacian template

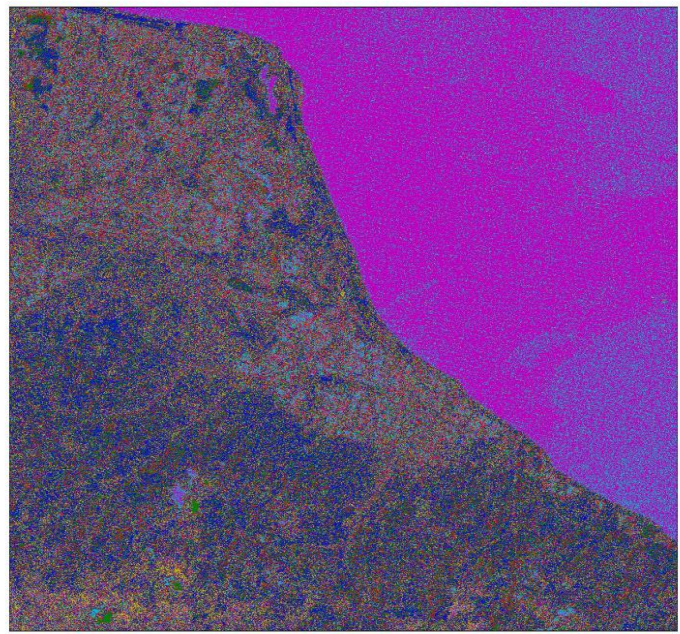

Fig. 16. Different elements in band 5, distinguished by Laplacian template

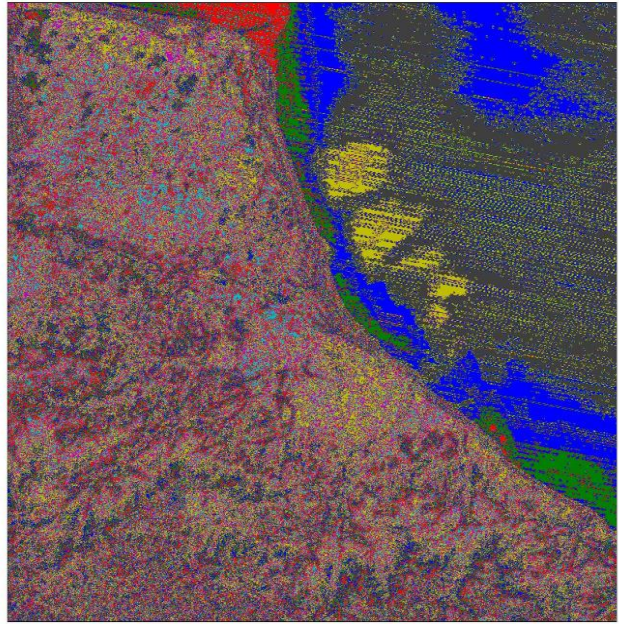

Fig. 17. Different elements in band 6, distinguished by Laplacian template

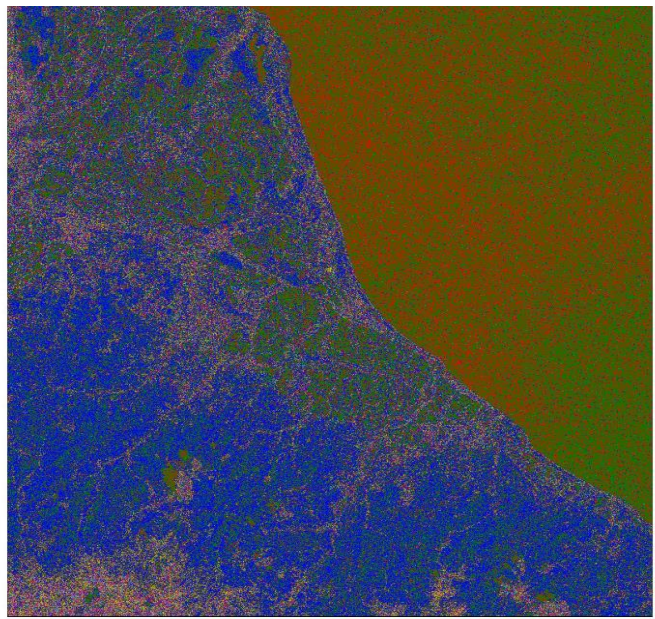

Fig. 18. Different elements in band 7, distinguished by Laplacian template

Comparison between the Fig. 4-10 and Fig. 12-18 reveals the fact that the proposed method in this paper is better in distinguishing various phenomena in the image. In the following figure, classification of the various phenomena in the image is given, using the method of parallelepiped classification, with the overall accuracy of 73 percent.

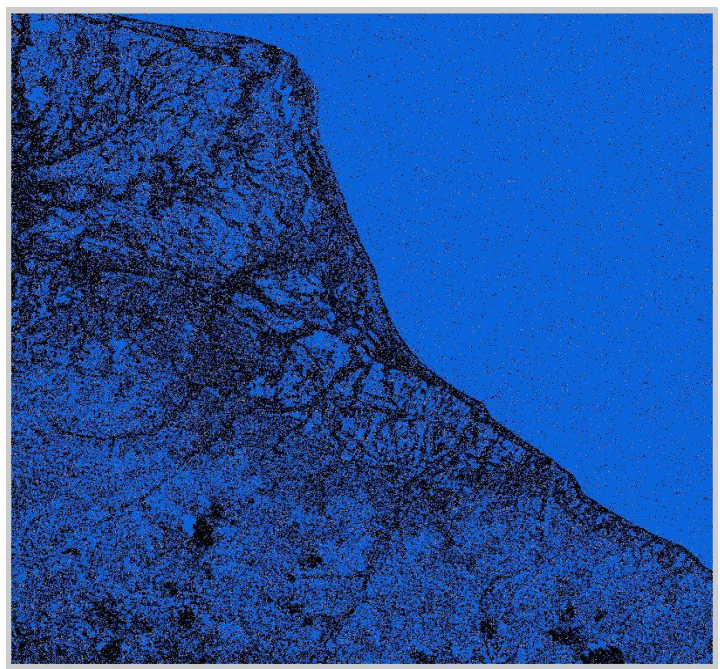

Fig. 19. Classified image in 6 bands (without the thermal band 6), using the image smoothing method images.

## 5. Conclusion

In this paper a new method of image smoothing is presented, which is based on the gradient norm minimization. The minimization problem is discretized and solved based on the finite difference method, and a symmetric $5 \times 5$ cell template is derived. Convolution of the derived template with an array of cells results in distinguishing the various phenomena that are present in the cells. An application of the obtained results is presented in the field of environmental remote sensing, using the Landsat multispectral satellite imagery. In this application, the Landsat image in 7 bands is smoothed with the derived template. Various elements in the picture, including Caspian Sea, jungles, and soils are distinguished and subsequently classified. Comparison between the results of smoothing of the derived template and the independent method of Laplacian template shows that the method proposed in this paper works better in discriminating the different elements that are present in the image.

The derived template is just one member of the large family of smoothing functions that can be derived using minimization problems. In order to do so, one needs to minimize higher order gradient and Laplacian operators, the so-called iterated Laplacian minimization. This was not pursued in this paper. However, the efficiency of the proposed method in this paper can be a motivation for researchers who work in these areas.

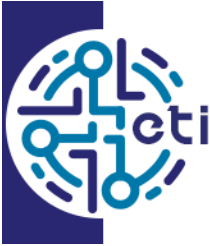
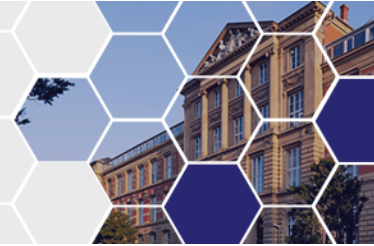